\documentclass[twocolumn, pra, aps, 10pt, longbibliography, floatfix]{revtex4-2}      
\usepackage{silence}        
\usepackage{graphicx,subfigure}                           
\usepackage[normalem]{ulem}                                
\usepackage[dvipsnames]{xcolor}                           
\usepackage{physics}                                      
\usepackage{dsfont}                                       
\usepackage{amsfonts, amssymb, amsmath}			          
\definecolor{mycolor}{rgb}{0.61, 0.11, 0.19}
\usepackage[
colorlinks=true, 
linkcolor=mycolor, 
citecolor=mycolor, 
urlcolor=mycolor]{hyperref}                                
\usepackage{orcidlink}           
\usepackage{times}
\usepackage{appendix}

\begin{document}
\title{Controlled dynamics of a multi-component discrete-time quantum walker}
\author{Vikash Mittal\,\orcidlink{0000-0002-4384-6992}}
\email[]{vikashmittal.iiser@gmail.com}
\affiliation{Institute of Physics, Polish Academy of Sciences, Aleja Lotnikow 32/46, PL-02668 Warsaw, Poland}     
\author{Tomasz Sowi\'nski\,\orcidlink{0000-0002-7970-4371}}
\email[]{tomasz.sowinski@ifpan.edu.pl}
\affiliation{Institute of Physics, Polish Academy of Sciences, Aleja Lotnikow 32/46, PL-02668 Warsaw, Poland} 

\begin{abstract}
We investigate a discrete-time quantum walk of a three-component particle on a one-dimensional lattice. The coin belongs to a three-parameter family of SU(3) operators generated by pairwise off-diagonal Gell–Mann couplings, enabling systematic tuning of mixing among the internal components. We map finite- and long-time transport over a wide parameter range using the position distribution, the standard-deviation growth rate, conditional left/right moments, side populations, and outer-window probabilities. The standard deviation is consistent with ballistic growth across the sampled parameter range, whereas the coefficient and the distribution's directional structure vary strongly with the couplings. The results identify extended parameter regions with suppressed spreading, central probability accumulation, and pronounced left–right asymmetry, demonstrating controllable reshaping of three-component quantum-walk transport.
\end{abstract}

\maketitle

\section{Introduction}
To develop sustainable quantum technologies and harness the power of quantum phenomena, precise control of quantum superposition, interference, and correlations is essential~\cite{Brif_2010_NJP, Nori_2012_IEEE}. Without such control, these features manifest as fluctuations and decoherence, severely limiting quantum-enhanced sensing, simulation, communication, and computation. In this context, quantum walks, the quantum analogs of classical random walks, provide a simple yet powerful framework for investigating controlled quantum dynamics~\cite{Aharonov1993, Nayak2000, Kempe2003, VenegasAndraca2012, Kadian2021}. Over the past two decades, they have found numerous applications in quantum information~\cite{Childs2004, Shenvi2003, Renato2013, Kadian2021, Mittal2025, Mittal2026, Chen2025, Sperling2024}, quantum computation~\cite{Childs2009, Childs2013, Singh2021}, and quantum transport~\cite{Jex_2020_PRA, Molfetta_2024_IOP}. In particular, quantum walks have become a key tool for understanding quantum control at the level of single~\cite{Kempe2003, Kitagawa2010, Kitagawa2012} and few particles~\cite{Goyal2010, Mittal2026a}. Further, they serve as building blocks for algorithmic speedups~\cite{Childs2003, Renato2013} and programmable simulators on complex graphs and lattices~\cite{Kitagawa2010, Qiang2024}.

In one spatial dimension, discrete-time coined quantum walks conventionally use a two-state coin to control the walker’s motion to the left or right. A natural extension that preserves one-dimensional geometry is to include a third coin state corresponding to “stay” (or “rest”) in addition to left and right movement. The resulting three-component (or “lazy”) quantum walk is the minimal coined walk capable of exhibiting intrinsic localization and a richer spectral structure~\cite{Konno_2005_PRE}. For the canonical three-state Grover coin, the evolution operator acquires a momentum-independent eigenvalue, leading to an exponentially localized stationary component around the initial position of the walker, coexisting with two ballistic wavefronts—a behavior with no counterpart in the homogeneous two-state walks~\cite{Konno_2005_PRE}. Along the same line of thought, subsequent works have developed detailed spectral and probabilistic theories for such three-state walks, primarily for specific choices of $3\times 3$ coins including Grover~\cite{Stefanak_2012_EPJD, Sadowski_2016_IOP}, Fourier~\cite{Konno_2017_arXiv}, and generalized Grover-type constructions~\cite{Saha_2021_IEEE}, and have classified families of coins that exhibit point spectrum (“trapping” coins), derived explicit formulas for trapping probabilities and peak group velocities, and proved limit theorems for the long-time position distribution~\cite{Stefanak_2012_EPJD, Kiumi_2022_JPA, Stefanak_2014_PRA, Machida_2014_arXiv, Machida_2015_PRA, Mandal_2022_PRA}. The effects of decoherence and noise on three-component walks have also been investigated~\cite{Tude_2022_PhysicaA, Tude_2022_QST, Kendon_2026_IOP}.

Most investigations into three-component quantum walks have therefore been motivated by localization and have often employed effective two-component recursions or specialized coin parameterizations. In parallel, studies of topological aspects of discrete-time quantum walks have largely concentrated on two-component split-step walks~\cite{Kitagawa2010, Asboth2012, Mittal2021} with only a few recent works addressing systems with three internal states~\cite{Ortwin_2026_PRR}. This leaves open the question of how more general SU$(3)$ coin constructions, designed with explicit control over pairwise couplings between coin basis states, can be used to systematically tune transport properties in three-component walks.

In this work, we address this gap by analyzing a discrete-time quantum walk of a three-level (qutrit) particle on a one-dimensional lattice. The system’s dynamics are determined by a conditional shift operation and a single SU$(3)$ rotation, realized as a weighted linear combination of three distinct Gell–Mann generators, each of which couples a different pair of internal coin states. This particular parametrization of the coin operator permits a controlled and systematic analysis of how the activation and continuous tuning of couplings to the third component influence the walker’s spreading rate, localization characteristics, and directional asymmetry in the resulting probability distribution. Rather than classifying localization or deriving a general weak limit~\cite{Stefanak2014, Segawa2016}, here we focus on how a physically transparent pair-coupling generator redistributes finite-time ballistic transport, conditional side widths, side displacements, and outer-front weight across a two-dimensional parameter window.


The rest of the article is organized as follows: it begins with an introduction to the underlying system, which has three components, and establishes the terminology for the rest of the article in Section~\ref{sec:framework}. We then focus on transport and quantify the dynamics by evolving the system for a longer time in Sec .~\ref{sec:ballistic}. Further, Sec.~\ref{sec:fixedtime} is dedicated to the analysis by fixing the time steps and traversing the full parameter space of the coin. In Sec.~\ref{sec:windows}, we present a side-resolved analysis to further quantify the dynamics with the role of parity discussed in Sec.~\ref{sec:parity}. Sec .~\ref {sec:control} examines transport relative to reference ballistic fronts. Finally, we conclude in Sec .~\ref {sec:conclusion} with a discussion on possible future extensions of the present work.

\section{The framework}
\label{sec:framework}
We consider a discrete-time quantum walk of a three-component particle (referred to as a ``walker'') on a one-dimensional lattice. The Hilbert space of the system is therefore the tensor product
\begin{equation}
    \mathrm{H} = \mathrm{H}_{\mathrm{C}}\otimes\mathrm{H}_{\mathrm{X}}
\end{equation}
of the three-dimensional Hilbert space $\mathrm{H}_{\mathrm{C}}$ related to the internal degrees of freedom, which is spanned by basis states $\{\ket{\mathtt{R}},\ket{\mathtt{G}},\ket{\mathtt{B}}\}$, and the $N$-dimensional Hilbert space $\mathrm{H}_{\mathrm{X}}$ related to spatial position, spanned by orthogonal states $\{\ket{x}\,|\,x\in\mathbb{Z}\}$. A single time step of the three-state quantum walk is governed by the following time evolution operator
\begin{equation}
    \mathcal{U}(\alpha,\beta,\gamma)
    = \mathcal{S}\,\mathcal{C}(\alpha,\beta,\gamma),
\end{equation}
where the operator $\mathcal{S}$ is the conditional translation of the walker of the form
\begin{equation}
    \begin{aligned}
    \mathcal{S} = \sum_{x\in\mathbb{Z}} \Big(
        &\ket{\mathtt{R}}\!\bra{\mathtt{R}}\otimes\ket{x+1}\!\bra{x}
      + \ket{\mathtt{G}}\!\bra{\mathtt{G}}\otimes\ket{x}\!\bra{x} \\
      &+ \ket{\mathtt{B}}\!\bra{\mathtt{B}}\otimes\ket{x-1}\!\bra{x}
    \Big).
    \end{aligned}
\end{equation}
It means that, by definition, at each time step, the $\ket{\mathtt{R}}$ component moves one site to the right, the $\ket{\mathtt{B}}$ component moves one site to the left, and the $\ket{\mathtt{G}}$ component remains at the same site.
On the other hand, the coin-toss operator $\mathcal{C}(\alpha,\beta,\gamma)$ acts non-trivially only in the coin subspace, and it mixes the different components of the coin. To make the analysis broad and comprehensive, we choose the coin operator from the $\mathrm{SU}(3)$ group that is generated by rotations with respect to the Gell--Mann matrices. This is a natural extension of the two-component system, allowing us to explore a larger region of the coin-parameter space. The most general $\mathrm{SU}(3)$ coin operator can be written as~\cite{Nielsen_Chuang_2010}
\begin{equation}
    \mathcal{C} = \exp\!\left[-i {\vb{\Theta}} \vdot \boldsymbol{\lambda} \right]\otimes\mathds{1}_{\mathrm{X}},
\end{equation}
where $\vb{\Theta} = [\Theta_1, \Theta_2, \dots, \Theta_8]$ is an eight-dimensional real vector of rotation angles and $\boldsymbol{\lambda} = [\lambda_1, \lambda_2, \dots, \lambda_8]$ is a vector containing the Gell--Mann matrices. In this work, we restrict the coin to the three-parameter SU$(3)$ family that allows us to separate and then couple the different internal components in a controlled way. For this purpose, we choose $\lambda_1, \lambda_4$, and $\lambda_6$ with $\Theta_1 = \beta$, $\Theta_4 = \alpha$, and $\Theta_6 = \gamma$, respectively. This results in the coin operator that we are going to use 
\begin{equation}
    \mathcal{C}(\alpha,\beta,\gamma)
    = e^{-i (\beta \lambda_1 + \alpha \lambda_4 + \gamma \lambda_6)}\otimes\mathds{1}_{\mathrm{X}},
    \label{eq:cointoss}
\end{equation}
where the relevant Gell--Mann matrices in use are
\begin{gather*}
    \lambda_1 =
    \begin{bmatrix}
        0 & 1 & 0 \\
        1 & 0 & 0 \\
        0 & 0 & 0
    \end{bmatrix},\quad
    \lambda_4 =
    \begin{bmatrix}
        0 & 0 & 1 \\
        0 & 0 & 0 \\
        1 & 0 & 0
    \end{bmatrix},\quad
    \lambda_6 =
    \begin{bmatrix}
        0 & 0 & 0 \\
        0 & 0 & 1 \\
        0 & 1 & 0
    \end{bmatrix}.
\end{gather*}
The motivation behind this choice of generators is straightforward. In the basis $\{\ket{\mathtt{R}},\ket{\mathtt{G}},\ket{\mathtt{B}}\}$, the generator $\lambda_4$ mixes the states in the $\{\ket{\mathtt{R}},\ket{\mathtt{B}}\}$ subspace, $\lambda_1$ couples $\{\ket{\mathtt{R}},\ket{\mathtt{G}}\}$, and $\lambda_6$ couples $\{\ket{\mathtt{G}},\ket{\mathtt{B}}\}$. The three rotation angles $(\alpha,\beta,\gamma)$ therefore provide direct control over the pairwise couplings between the three internal components. Moreover, this parametrization admits a simple decoupling limit. For $\beta = \gamma = 0$, the coin reduces to a rotation within the $\{\ket{\mathtt{R}},\ket{\mathtt{B}}\}$ subspace, while $\ket{\mathtt{G}}$ is left unchanged. Therefore, the evolution operator $\mathcal{U}(\alpha,0,0)$ reduces to the block diagonal form with one block controlling the dynamics of the two components in the subspace $\{\ket{\mathtt{R}},\ket{\mathtt{B}}\}$ and the second is associated with a trivial evolution of the $\ket{\mathtt{G}}$ component. This limit will serve as a useful reference for the dynamics with various couplings turned on.

At any discrete time $t$, a quantum state of the walker can be written as
\begin{equation}
    \begin{aligned}
            \ket{\Psi(t)}
            =&\sum_{x\in\mathbb{Z}} \Big(
                \psi_{x,\mathtt{R}} (t)\ket{\mathtt{R}}
              + \psi_{x,\mathtt{G}}(t) \ket{\mathtt{G}}
              + \psi_{x,\mathtt{B}}(t) \ket{\mathtt{B}}
            \Big)\otimes\ket{x},
    \end{aligned}
\end{equation}
and the corresponding probability distribution of finding the walker at lattice site $x$ and at time $t$ reads
\begin{equation}
    P(x, t) = \sum_{\sigma = \{\mathtt{R}, \mathtt{G}, \mathtt{B}\}} \abs{\psi_{x, \sigma}(t)}^2.
\end{equation}
Using this probability distribution, we can straightforwardly evaluate the expectation value of any function of the position $f(x)$, as
\begin{equation}
    \expval{f(x)}_t := \sum_x f(x) P(x,t).
\end{equation}

To make the analysis as clear as possible, we initialize the walker in the quantum state of the walk in the form
\begin{equation}
    \ket{\Psi(0)} = \dfrac{1}{\sqrt{2}} (\ket{\mathtt{R}} + \ket{\mathtt{B}}) \otimes \ket{0},
    \label{eq:initialstate}
\end{equation}
{\it i.e.}, the walker is perfectly localized at a single lattice site, and the internal state of the coin has no component along $\ket{\mathtt{G}}$. This choice enables us to explicitly quantify the effect of the various couplings, which result in the transfer of population from the states $\ket{\mathtt{R}}$ and $\ket{\mathtt{B}}$ to $\ket{\mathtt{G}}$. For the initial state $\ket{\Psi(0)}$, given by Eq.~\eqref{eq:initialstate}, we evolve the state for $t$ time steps as
\begin{equation}
     \ket{\Psi(t)} = \mathcal{U}(\alpha, \beta, \gamma)^t \ket{\Psi(0)}
\end{equation}
for various combinations of couplings. We fix $\alpha=\pi/4$ throughout the whole article, to ensure that the coin operator has equal magnitude of transition amplitudes between $\{ \ket{\mathtt{R}}, \ket{\mathtt{B}}\}$ components, and we systematically scan the remaining couplings \(\beta\) and \(\gamma\). 
\begin{figure}
    \centering
    \includegraphics[width=0.5\textwidth]{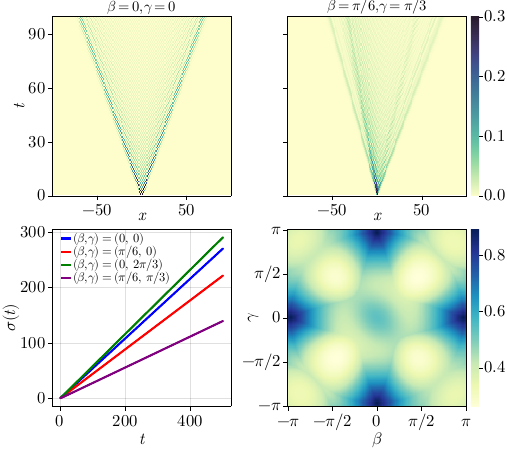}
    \caption{(Top Row): The probability distribution $P(x,t)$ for the three–state walk for two representative choices of the SU$(3)$ coin parameters $(\beta,\gamma)$, as indicated in each panel. The walker is initially localized at the origin with the coin in equal superposition of $\ket{\mathtt{R}}$ and $\ket{\mathtt{B}}$ as given by Eq.~\eqref{eq:initialstate}. The color scale shows the total probability at each lattice site. The red and the blue lines indicate the conditional side-resolved mean positions $X_L$ and $X_R$ (defined in Sec.~\ref{sec:windows}). (Bottom Left): The time evolution of \(\sigma(t)\) for four representative pairs of couplings $(\beta,\gamma)$. (Bottom Right): slope $m(\beta,\gamma)$ that is calculated from the long-time evolution over the full range of couplings $\beta$ and $\gamma$.}
    \label{fig:dynamics}
\end{figure}

In the top row of Fig.~\ref{fig:dynamics}, we plot the spatio–temporal density $P(x,t)$ for $(\beta,\gamma) = (0,0)$, and $(\pi/6,\pi/3)$, with the color scale indicating the local probability distribution. The dynamics display two dominant ballistic fronts that propagate approximately linearly in time, forming the familiar lightcone structure known from conventional two–component quantum walks~\cite{Kempe2003}. The inner structure of the probability distribution, however, depends sensitively on the couplings $\beta$ and $\gamma$ with the third component of the coin. For $(\beta,\gamma) = (0,0)$ the third component is decoupled and the evolution reduces to a standard two–component walk in the $\{\ket{\mathtt{R}},\ket{\mathtt{B}}\}$ subspace and the probability is concentrated in two sharp ballistic peaks, with only weak interference inside the lightcone. As $\beta$ and $\gamma$ are switched on, the interference effects with the third components enter the dynamics, which redistributes the weight from the ballistic fronts towards the interior of the cone, and the density profile near the origin becomes visibly broader. The case $(\beta,\gamma) = (\pi/6,\pi/3)$ shows mixing, with the ballistic peaks noticeably attenuated and a larger fraction of the probability remaining near the initial site for long times, indicating slower effective spreading.

\section{Ballistic transport}
\label{sec:ballistic}
A more quantitative measure of the spreading is provided by the standard deviation of the position distribution, which is defined as
\begin{equation}
    \sigma(t)
    = \left[\langle x^2\rangle_t - \langle x\rangle_t^2\right]^{1/2}.
    \label{eq:variance}
\end{equation}
In the bottom left panel of Fig.~\ref{fig:dynamics}, we plot $\sigma(t)$ as a function of time for four representative choices of the coin parameters $(\beta,\gamma)$. For all four parameter sets, the curves are straight lines over the entire time window shown, demonstrating that the dynamics remain ballistic even when the coupling to the third internal component is switched. We numerically verified that transport is ballistic for any set of parameters $(\beta,\gamma)$. The different curves are distinguished only by their slopes, and a particular combination of $(\beta,\gamma)$ can enhance as well as suppress them. To quantify this effect across the full parameter space, we evolve the system for a long enough time, $t = 2000$, and fit the linear dependence, $\sigma(t) = m(\beta,\gamma)\cdot t$. To ensure the system has settled and is sufficiently stable, we use data from $t = 500$ to $t = 2000$ to estimate the linear coefficient $m(\beta,\gamma)$. The linear coefficient $m(\beta,\gamma)$ sets the scale of the lightcone radius at a given time and sets the asymptotic scale of the standard deviation.

The dependence of $m(\beta,\gamma)$ on the coin parameters is shown in the bottom right panel of Fig.~\ref{fig:dynamics}. It displays dark regions (corresponding to large $m$) where the walker undergoes rapid ballistic expansion, and bright regions (corresponding to small $m$) where the spreading is strongly reduced, although it remains linear in time. The minima of $m(\beta,\gamma)$ form extended curves rather than isolated points, indicating that the suppression of spreading is robust over finite regions of parameter space and does not rely on fine-tuning. In addition to the bright and dark regions, we observe smoothly varying boundary structures that form a circle of radius $\pi$. However, these boundaries are not accompanied by any significant transitions in the function $m(\beta, \gamma)$.
\begin{figure*}
    \centering
    \includegraphics[width=\textwidth]{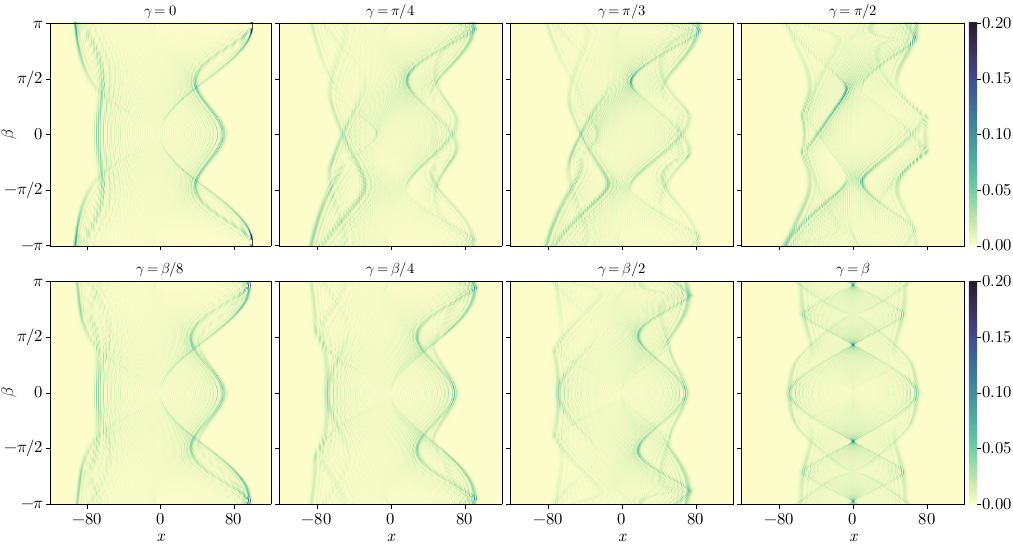}
    \caption{The probability $P(x,t)$ of finding the walker at lattice site $x$ at fixed time $t = 100$ and different coupling angle $\beta$ going from $-\pi$ to $\pi$. (Top Row): fixed values of $\gamma = 0, \pi/4, \pi/3$ and $\pi/2$. (Bottom Row): fixed ratios $\gamma/\beta = 1/8, 1/4, 1/2$, and $1$. The dynamics is symmetric under the swap $\beta \rightarrow \gamma$, except for the mirror reflection about $x = 0$.}
    \label{fig:fixtime}
\end{figure*}

\section{Fixed-time analysis}
\label{sec:fixedtime}
To resolve more clearly how the couplings $(\beta,\gamma)$ control the spatial structure of the wave function, and hence the probability distribution, we fix the total evolution time, $t = 100$, and examine the dependence of $P(x,t)$ on the different couplings coming from the coin parameters. In the top row of Fig.~\ref{fig:fixtime}, we first scan the coupling angle $\beta$ for some fixed $\gamma$, and we plot the distribution $P(x,t)$. We choose $\gamma = 0, \pi/4, \pi/3$ and $\pi/2$. Each horizontal cut corresponds to the spatial profile at a given $\beta$. As $\beta$ moves away from zero, the ballistic fronts shift, their amplitudes change, and the interference fringes become more intricate. We observe regions of parameter space where the ballistic peaks ramify into multiple additional peaks, yielding more intricate patterns. We note a strong finite-time concentration of the probability for $\gamma = 0$ around $\beta = -\pi$ and $\pi$ on the right side of the initial position of the walker. Apart from this, we do not observe a significant localization across the lattice for any combination of the parameters.

Complementary information is obtained further by not explicitly fixing $\gamma$ and making it comparable to $\beta$. In the bottom row of Fig.~\ref{fig:fixtime}, we again plot the $P(x,t)$ as a function of $x$ and $\beta$ but now for $\gamma/\beta = 1/8, 1/4, 1/2$, and $1$. Comparing the two cases reveals how introducing $\gamma$ in a different way modifies the $\beta$–dependence of the ballistic peaks and interference pattern. One contrasting feature of this case is that the distribution remains symmetric about $\beta = 0$, and the probability near the central region remains significantly low. For $\gamma/\beta = 1/8$, the ballistic cones remain relatively symmetric, and the main effect of changing $\beta$ is to shift their positions and modulate their amplitudes, as in the case $\gamma = 0$. For $\gamma = \beta$, the dynamics restores the symmetry $x \leftrightarrow -x$. The ballistic fronts bend strongly, and in the vicinity of $\beta \simeq \pm\pi/2 $ and $ \pm \pi$, they intersect and form enhanced probability ridges that connect the central region to the edges and lead to significant amplification of the probability near the initial location. 

These features identify parameter combinations $(\beta, \gamma)$ where the coupling between the two–component subspace and the third component leads to strong hybridization of bulk modes, and they foreshadow the more global structures that appear when we consider spatially integrated observables over different regions of the lattice. 

\section{Side-Resolved analysis}
\label{sec:windows}
\begin{figure}
    \subfigure{\includegraphics[width=0.50\textwidth]{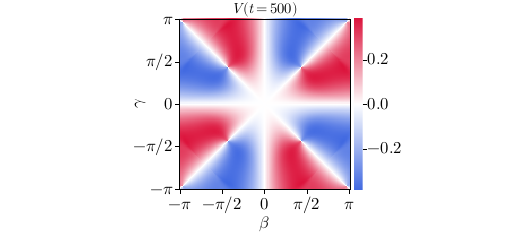}}
    \subfigure{\includegraphics[width=0.50\textwidth]{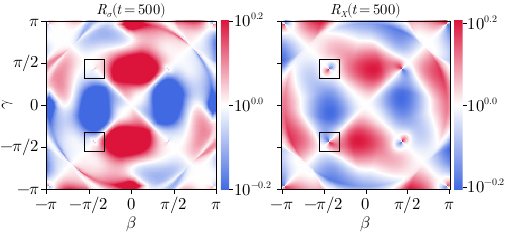}}
    \caption{Top Row: The side-population imbalance $V(t)$, defined in Eq.~\eqref{eq:imbalance}. Bottom Row: Ratio of variances $R_{\sigma} =  \sigma_{\mathrm{R}}^2/\sigma_{\mathrm{L}}^2$ (Left) and the mean-position $R_X = -X_R/X_L$ (Right) between the right and left halves of the lattice as a function of the coin parameters $(\beta,\gamma)$. All the plots use a fixed evolution time $t = 500$. The decoupling limit $\beta=\gamma=0$ is nearly symmetric, while finite couplings generate extended regions of pronounced directional bias, organized along smooth curves in parameter space.}
    \label{fig:ratios}
\end{figure}

To provide a more complete characterization of the dynamics, we employ a side-resolved analysis in which relevant observables are evaluated separately on the two sides of the initially occupied site, thereby quantifying potential left–right asymmetries in transport. We first introduce the side-restricted density distributions
\begin{subequations}
    \begin{align}
      p_{\mathrm{L}}(x,t) = \begin{cases}
          P(x,t), \;\;\; &x < 0 \\
          0, \;\;\; &x \ge 0 
      \end{cases}\\
      p_{\mathrm{R}}(x,t) = \begin{cases}
          P(x,t), \;\;\; &x > 0 \\
          0, \;\;\; &x \le 0.
      \end{cases}
    \end{align}
\end{subequations}
From these distributions, we define side-resolved populations
\begin{subequations}
    \begin{align}
        P_{\mathrm{L}}(t)&= \sum_x p_{\mathrm{L}}(x,t), \\
        P_{\mathrm{R}}(t)&= \sum_x p_{\mathrm{R}}(x,t),
    \end{align}
\end{subequations}
side-resolved mean positions
\begin{subequations}
    \begin{align}
        X_{\mathrm{L}}(t)&= \dfrac{1}{P_{\mathrm{L}}(t)} \sum_x x\, p_{\mathrm{L}}(x,t), \\
        X_{\mathrm{R}}(t)&= \dfrac{1}{P_{\mathrm{R}}(t)} \sum_x x\, p_{\mathrm{R}}(x,t),
    \end{align}
\end{subequations}
and the side-resolved variances
\begin{subequations}
    \begin{align}
        \sigma_{\mathrm{L}}^2(t)
        = &\dfrac{1}{P_{\mathrm{L}}(t)}\sum_{x}\big(x- X_{\mathrm{L}}\big)^2 p_{\mathrm{L}}(x,t),
        \\
        \sigma_{\mathrm{R}}^2(t)
        = &\dfrac{1}{P_{\mathrm{R}}(t)}\sum_{x} \big(x- X_{\mathrm{R}}\big)^2 p_{\mathrm{R}}(x,t).
    \end{align}
\end{subequations}
We emphasize that the site $x=0$ is excluded by construction; consequently, the above side-resolved quantities involve only the portion of the distribution that has moved away from the initially occupied site. Using them, we introduce a measure of population imbalance (analogous to a visibility) as
\begin{equation}
    V(t) = \dfrac{P_{\mathrm{R}}(t) - P_{\mathrm{L}}(t)}{P_{\mathrm{R}}(t) + P_{\mathrm{L}}(t)}
    \label{eq:imbalance}
\end{equation}
and the dimensionless ratios
\begin{equation}
    R_{\sigma} = \dfrac{\sigma_R^2}{\sigma_L^2},\;\;\; R_X = \dfrac{X_R}{\abs{X_L}}.
\end{equation}
We plot these quantities in Fig.~\ref{fig:ratios} with the side-population imbalance $V$ in the top row, and $R_\sigma$ and $R_X$ in the bottom row, as a function of the coin parameters $(\beta,\gamma)$ for a fixed time step $t=500$. This time step is long enough to ensure the saturation of the ratios. Inspecting the plot of $V$ indicates that, when only one of the couplings ($\beta$ or $\gamma$) is introduced, the population remains confined to the respective sides and the resulting variations are predominantly local. This behavior is evidenced by the white regions along the lines $\beta = 0$ and $\gamma = 0$. Further, to analyze the ratios, we use the logarithmic scale in the plot because we are interested in relative rather than absolute magnitudes.  The values close to unity indicate nearly symmetric transport, whereas values significantly larger (or smaller) than one signal a directional asymmetry. We again recognize an emergence of a circular boundary familiar from the density plot of $m(\beta, \gamma)$ in Fig.~\ref {fig:dynamics}. We clearly see extended red and blue regions separated by well-defined boundaries, indicating a continuous transfer of probability between the two sides as we sweep through various parameter regions.  The variance ratio $R_\sigma$ (bottom left panel of Fig.~\ref{fig:ratios}) highlights that these anisotropic regimes are organized along smooth curves in the $(\beta,\gamma)$ plane, forming lobes where right- and left-side spreading dominate.  

The complementary picture is provided by the corresponding analysis of the mean-position ratio $R_X$ (on the bottom right side of Fig.~\ref{fig:ratios}). The behavioral pattern across the parameter space is more involved than the one in $R_\sigma$. Apart from a smooth circular boundary, we have additional continuous boundaries that mark crossovers in the probability density from left to right and vice versa. The global map reveals that the mean-position asymmetry organizes along smooth curves similar to those seen in the variance ratio, but with slightly different relative weights, reflecting the different sensitivity of the mean and the variance to the detailed shape of the probability distribution.

The fact that the same kind of boundary structure appears in both $R_\sigma$ and $R_X$ shows that changes in the typical width and typical displacement are tightly correlated in the walk with a parametrized SU$(3)$ coin. Whenever the dominant side of the spreading switches, both the variance ratio and the mean-position ratio cross unity in a coordinated manner.

Another prominent feature, clearly visible in $R_X$, is the structure observed in the vicinity of the parameter values $\beta = \gamma \approx \pm \pi/2$. We observe an enhanced sensitivity of the side-resolved ratios. To examine this behavior in greater detail, we focus on the vicinity of $\beta \sim -\pi/2$ and $\gamma \sim \pm \pi/2$, as marked by the squares in the bottom row of Fig.~\ref{fig:ratios}. For parameter values near $\gamma \sim -\pi/2$ (square at the bottom), there is a marked increase in the probability within the two halves of the lattice, which connects smoothly to the local behavior in that region. In contrast, a closer inspection of the ratios near $\gamma \sim \pi/2$ (square on the top) reveals an additional crossover of the probability from one side of the lattice to the other, superimposed on the enhanced sensitivity. This additional crossover exhibits a complementary relationship between $R_\sigma$ and $R_X$, {\it i.e.}, whenever a crossover is observed in $R_\sigma$, it is absent in $R_X$, and vice versa. We also observe these points at the diagonal and anti-diagonal positions near $\beta \sim \pm \pi$ and $\gamma \sim \pm \pi$, and they appear not to follow a periodic pattern.
\begin{figure}
     \subfigure{\includegraphics[width=0.50\textwidth]{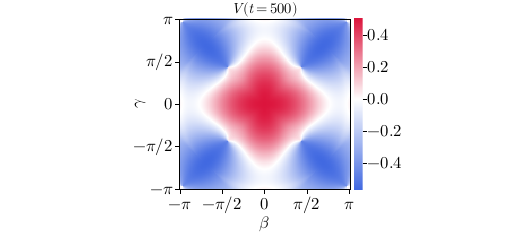}}
    \subfigure{\includegraphics[width=0.50\textwidth]{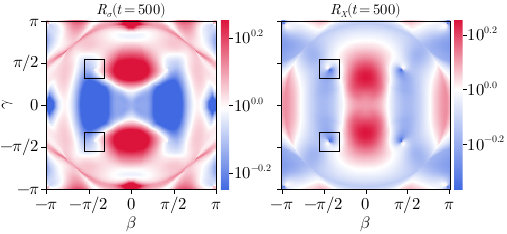}}
    \caption{Population imbalance, ratio of variances $R_{\sigma} =  \sigma_{\mathrm{R}}^2/\sigma_{\mathrm{L}}^2$ (Left) and the mean-position $R_X = -X_R/X_L$ (Right) between the right and left halves of the lattice at fixed evolution time $t = 500$ as a function of the coin parameters $(\beta,\gamma)$ for a state given in Eq.~\eqref{eq:initialstate2}}
    \label{fig:ratios2}
\end{figure}

\section{Role of the parity}
\label{sec:parity}

So far, our analysis has focused on the initial coin state introduced in
Eq.~\eqref{eq:initialstate}. This state is an eigenstate of the parity
operator 
\begin{equation}
    \mathcal{P}
    =
    \begin{bmatrix}
        0 & 0 & 1 \\
        0 & 1 & 0 \\
        1 & 0 & 0
    \end{bmatrix}
    \otimes
    \sum_{x\in\mathbb{Z}}\ket{-x}\bra{x},
\end{equation}
which combines the exchange of the right- and left-moving
coin components with spatial reflection. The evolution operator
satisfies
\begin{equation}
    \mathcal{P}\mathcal{U}(\alpha,\beta,\gamma)\mathcal{P}
    =
    \mathcal{U}(\alpha,\gamma,\beta).
\end{equation}
Thus, for $\beta=\gamma$, parity is conserved, and the original
initial state produces a reflection-symmetric probability
distribution, $P(x,t)=P(-x,t)$. Unequal couplings generally break
this symmetry of the evolution and allow asymmetric transport.

To investigate the dependence of the transport structures on
the initial preparation, we instead consider the walker in the state
\begin{equation}
    \ket{\Psi(0)}
    =
    \frac{1}{\sqrt{2}}
    \left(
        \ket{\mathtt{R}}+i\ket{\mathtt{B}}
    \right)\otimes\ket{0}.
    \label{eq:initialstate2}
\end{equation} 
Evidently, this state is not an eigenstate of the parity operator defined above,
so its evolution is not constrained to produce a symmetric
probability distribution, even when $\beta=\gamma$ and, indeed, it already produces asymmetric motion in the decoupled limit $\beta=\gamma=0$. Please note that this lack of definite parity can also be achieved by initializing the walker in an asymmetric superposition of several lattice sites. Comparing these two preparations allows us to assess whether the previously identified transport structures persist when the parity constraint on the initial
state is removed.

In Fig.~\ref{fig:ratios2} we plot the corresponding side-resolved diagnostics at \(t=500\). Despite explicitly changing the parity character of the initial coin
state, the parameter maps retain the extended regions in which \(R_{\sigma}\) and \(R_X\) differ substantially from unity. The reappearance of these
extended structures for a parity-breaking initial state shows that they
are not artifacts of the original parity-eigenstate preparation.
Instead, they reflect a persistent property of the interference and
mixing induced by the three-component coin.

We again mark the same regions which we marked in Fig.~\ref{fig:ratios} where the ratios vary much more rapidly than in the surrounding parameter
space using two rectangles in the bottom row of Fig.~\ref{fig:ratios2}. In both windows, a sharply localized feature
appears in \(R_{\sigma}\), accompanied by a strongly structured response
in \(R_X\). The latter has an approximately dipolar profile: parameter
values producing an enhanced right-resolved displacement lie immediately
adjacent to values producing a strongly suppressed ratio. Thus, a small
change in either \(\beta\) or \(\gamma\) near these points can generate a
comparatively large change in both the magnitude and direction of the
transport imbalance.

The survival of these anomalous high-sensitivity points is particularly
noteworthy. Their positions and detailed profiles can depend on the
initial-state phase, since this phase changes the interference between
the initially occupied coin components. Nevertheless, their
reappearance demonstrates that the mechanism responsible for the sharp
response is not eliminated when the initial state ceases to be a parity
eigenstate. The extended regions of side-resolved asymmetry and the localized
high-sensitivity structures should therefore be regarded as inherent
features of the considered family of three-component walks, rather than
as consequences of a single fine-tuned initial condition.

Taken together, the side-resolved visibility and the mean and variance ratios provide a compact, quantitative characterization of the directional transport induced by the SU$(3)$ coin. These observables refine the picture by showing how far and how widely the wave packet spreads on each side. In the decoupling limit $\beta=\gamma=0$, all diagnostics point to nearly symmetric, purely ballistic transport.

\section{Control with third-component}
\label{sec:control}
To complement the ratios \(R_{\sigma}\) and \(R_X\), we introduce a fixed-window measure relative to the front positions of a reference evolution. As a reference, we consider the conventional two-component walk obtained for \(\alpha=\pi/4\) and \(\beta=\gamma=0\), for which the third component decouples from the two dynamically active coin components. Starting from the symmetric coin state
\((\ket{\mathtt{R}}+\ket{\mathtt{B}})/\sqrt{2}\), and the walker localized at the center, the two dominant ballistic peaks at \(t=100\) are located at
\(x_{\mathrm L}^{(0)}=-68\) and \(x_{\mathrm R}^{(0)}=68\). These
positions are used as fixed reference-front thresholds throughout the
\((\beta,\gamma)\) parameter scan. We define the left outer window by
\(x\leq x_{\mathrm L}^{(0)}\) and the right outer window by
\(x\geq x_{\mathrm R}^{(0)}\). This enables us to calculate the probability of finding the walker outside the cone of the standard two-component walk, using the quantities
\begin{align}
P_{\mathrm L}^{\mathrm{out}}(t)
&=
\sum_{x \le x_{\mathrm L}^{(0)}}
P(x,t),\\
P_{\mathrm R}^{\mathrm{out}}(t)
&=
\sum_{x \ge x_{\mathrm R}^{(0)}}
P(x,t).
\end{align}
These observables quantify the amount of probability that reaches or
extends beyond the left and right fronts of the reference walk. Since the boundaries are kept fixed while the coin parameters are varied,
changes in \(P_{\mathrm L}^{\mathrm{out}}\) and
\(P_{\mathrm R}^{\mathrm{out}}\) directly reveals how the coupling to the third component redistributes probability relative to the conventional ballistic scale.
\begin{figure}
    \centering
    \includegraphics[width=0.5\textwidth]{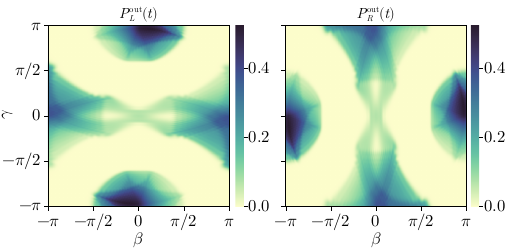}
    \caption{Total probability contained in the left (left panel) and right
    (right panel) outer windows at \(t=100\), plotted as a function of
    the coin parameters \((\beta,\gamma)\). The window boundaries are
    fixed by the positions of the two ballistic front peaks of the
    reference walk with \(\alpha=\pi/4\) and
    \(\beta=\gamma=0\).}
    \label{fig:window_probability}
\end{figure}    
The resulting parameter maps are shown in
Fig.~\ref{fig:window_probability}. Both outer-window probabilities vary
strongly and nonuniformly across the \((\beta,\gamma)\) plane. Extended
regions with appreciable \(P_{\mathrm L}^{\mathrm{out}}\) or
\(P_{\mathrm R}^{\mathrm{out}}\) demonstrate that the probability
transported to the outer part of the lattice can be enhanced by tuning
the couplings to the third component. Moreover, the left- and
right-window maps have clearly complementary structures. Thus, the same
three-component coupling that modifies the propagation range also
controls the direction in which the additional outer probability is
transferred.

This behavior provides a direct manifestation of transport control.
When \(\beta=\gamma=0\), the third component is decoupled, and the
reference front is recovered. Once the additional couplings are
turned on, interference involving the third component changes the
effective propagation of the wave packet. By selecting
\((\beta,\gamma)\), one can enhance the probability near and beyond the
reference front on the left, on the right, or suppress it on both sides.
The third coin component, therefore, acts as an additional control channel for shaping both the spatial extent and the directional character of the quantum walk. The enhancement of the outer-window probabilities demonstrates that the third-component couplings can transfer a controllable fraction of the
wave packet to spatial regions at or beyond the ballistic fronts of the
reference two-component walk. These observations show that the SU$(3)$ coin induces a highly structured and nontrivial transport regime in parameter space: a network of transport boundaries along which the ballistic front speed, the probability distribution across windows, and the left–right symmetry all undergo coordinated changes.

\section{Discussion \& Conclusion}
\label{sec:conclusion}
The parameter space explored in the present article represents only a restricted section of
the complete parameter space available to a three-component quantum
walk. A general \(SU(3)\) coin operator contains eight independent real
parameters, corresponding to the eight Gell-Mann matrices. In the present model, we restrict the coin to a
three-parameter family described by \((\alpha,\beta,\gamma)\), fix
\(\alpha\), and scan only the two-dimensional \((\beta,\gamma)\) plane.
The initial coin state introduces additional freedom: a general normalized
pure state in \(\mathbb{C}^{3}\), after removing its physically irrelevant
global phase, requires four independent real parameters. Consequently,
the most general combination of an \(SU(3)\) coin and a pure
three-component initial state spans a twelve-parameter space. Our
calculations probe a deliberately constrained subset of this space:
two coin parameters are varied, one coin angle is fixed, and the initial
coin amplitudes are held fixed. To examine the dependence on the state
preparation without undertaking an exhaustive multidimensional scan, we
supplement the original parity-eigenstate preparation with
Eq.~\eqref{eq:initialstate2}, which introduces a nontrivial relative phase
between the populated coin components. The recovery of both the extended regions of side-resolved asymmetry and the localized anomalous points for this second preparation indicates that these structures are not confined to the original parity-symmetric initial condition.

Additionally, the motivation for choosing this coin family was to establish a reference in the natural decoupling limit, in which one internal state acts as a localized spectator while the remaining two components undergo a conventional ballistic walk. It then enabled a systematic tuning of pairwise couplings between all three coin components.

For all sampled points in the chosen \((\beta,\gamma)\) window, the standard deviation is consistent with ballistic growth over the simulated long-time interval, while the fitted spreading coefficient varies substantially. Conditional side-resolved moments and side populations reveal extended regions of left–right asymmetry. These results concern finite-time redistribution and ballistic transport control; spectral localization and a full classification of the associated Floquet bands remain separate questions and potential future extensions of the present work.

These findings demonstrate that even in the absence of spatial disorder, adding a third internal state and an SU$(3)$ coin can substantially enrich the transport behavior of one-dimensional quantum walks and provide additional control knobs for engineering spreading, localization, and directional bias. The diagnostics developed here—variance slopes, and side-resolved ratios—are general and can be applied to any generic coin constructions and initial conditions. An interesting direction for future work is to relate the observed transport regimes more directly to the spectral properties of the underlying Floquet operator, including band structures and flat-band conditions, and to explore the consequences of spatially inhomogeneous or time-dependent coin parameters.

\section*{Data availability statement}
All numerical data presented in this paper are available online~\cite{zenodo}.

\section{Acknowledgments}
This research was supported by the National Science Centre (NCN, Poland) within the OPUS project No. 2023/49/B/ST2/03744. For the purpose of Open Access, the authors have applied a CC-BY public copyright licence to any Author Accepted Manuscript version arising from this submission.

\bibliography{zero.bib}

\end{document}